\documentclass[twocolumn,showpacs,preprintnumbers,amsmath,amssymb]{revtex4}

\usepackage{graphicx}
\usepackage{dcolumn}
\usepackage{bm}

\begin{document}


\title{Improved basis selection for the Projected Configuration Interaction method applied to heavy nuclei}

\author{Zao-Chun Gao$^{1,2}$}%
\author{Mihai Horoi$^1$}
\author{Y. S. Chen$^2$}
\affiliation{
$^1$Department of Physics, Central Michigan University, Mount
Pleasant, Michigan 48859, USA\\
$^2$China Institute of Atomic Energy P.O. Box 275-18, Beijing 102413, China
}

\date{\today}

\begin{abstract}
In a previous paper we proposed a  Projected Configuration Interaction method
that uses sets of axially deformed single particle states to build up the many body basis.
We show that the choice of the basis set is essential for the efficiency of the method,
and we propose a newly improved algorithm of selecting the projected basis states.
We also extend our method to model spaces that can accomodate both parities, and can include
 odd-multipole terms in the effective interaction, such as the octupole contributions.
Examples of $^{52}$Fe, $^{56}$Ni, $^{68}$Se, $^{70}$Se and $^{76}$Se are calcualted 
showing good agreement with the full Configuration Interaction  results. 
\end{abstract}

\pacs{21.60.Cs,21.60.Ev,21.10.-k}
\maketitle

\section{Introduction}

The full configuration interaction (CI) method \cite{si28,smrmp} 
using a spherical single particle (s.p.) basis and realistic Hamiltonians, also know as the nuclear shell model, has been very successful in describing various properties 
of the low-lying states in light and medium nuclei. The main limitations of this method are the exploding
dimensions with the increase of the number of valence nucleons, or/and with the increase of the
valence space. Although, there are continuous improvements to the CI codes \cite{ant,nushellx} and
computational resources, the exploding CI dimensions significantly restrict the ability to investigate 
heavy nuclei, especially those which exhibit strong collectivity.
The deformed mean-field approaches, however, have the ability to incorporate the collective effects
at the single particle level.
The mean-field description in the intrinsic frame naturally takes advantage of the spontaneous
symmetry breaking. This approach provides some physical insight, but the loss of
good angular momentum of the mean-field wave functions makes the comparison with the 
experimental data difficult.
The CI calculations in spherical basis provide the description in the laboratory frame.
The angular momentum is conserved, but the physical insight associated with the existence 
of an intrinsic state is lost. 
One important aspect of the CI approach is its ability of using all components of effective interactions
compatible with a given symmetry, but restricted to a chosen valence space. Examples of realistic Hamiltonians, 
such as the USD \cite{Wild84,si28}in the $sd$ shell, 
the KB3 \cite{Poves81}, FPD6 \cite{Richter91} and GXPF1 \cite{Honma02} in the $pf$ shell, 
have provided a very good base to study various nuclear structure problems microscopically.

The recent history of projection techniques combined with CI particle-hole configurations includes the
Projected Shell Model(PSM) \cite{Hara95,Chen01} and the Deformed Shell Model (DSM) proposed in 
reference \cite{dsm}. PSM uses a deformed intrinsic Nilsson+BCS basis
projected onto good angular momentum, and a multipole-multipole Hamiltonian that diagonalized 
in the space spanned by the projected states. The Nilsson model \cite{Nilsson55} has been proven 
to be very successful in describing the deformed intrinsic single particle states, and 
the quadrupole force was found to be essential for describing the rotational motion \cite{Elliott58}. 
PSM was proven to be a very efficient method in analyzing the phenomena
associated with the rotational states, especially the high spin states, not only for axial quadrupole deformation, 
but also for the octupole \cite{Chen01} and triaxial shapes \cite{Sheikh99,Gao06}.
However, its predictive power is limited because the mulitpole-multipole plus pairing
Hamiltonian has to be tuned to a specific class of states, rather than an region of the nuclear chart. 
The recently proposed DSM is using the same realistic effective Hamiltonian as the full CI method,
and a Hartree-Fock procedure to select the deformed basis. One can only assume that this procedure
would not be very accurate for quasi-spherical nuclei. The main difficulties for all these models is
a proper selection of the deformed basis. Their accuracy can only be assessed by comparison with
the exact results provided by the full CI method using the same effective Hamiltonian, and not by direct
comparison with the experimental data.
Other model using similar techniques includes  
MONSTER, the family of VAMPIRs \cite{Schmid04}, and
the Quantum Monte Carlo Diagonalization (QMCD) method \cite{Honma96}.

In a previous paper \cite{pci1} we proposed a new method of calculating the low-lying states
in heavy nuclei using many particle-hole configurations of spin-projected Slater determinants built on
multiple sets of deformed single particle orbitals. This Projected Configuration Interaction (PCI) method
takes advantage of inherent mixing induced by the projected Slater determinants of varying deformations
with the many particle-hole mixing typical for the Configuration Interaction (CI) techniques.
Direct comaparison between PCI and CI results are always possible, provided that the deformed s.p. states 
are always obtained starting from a valence space of shperical orbitals.
In Ref. \cite{pci1} we use a simple mechanism of selecting a number of basic deformed Slater determinants
in the $sd$ and $pf$ model space, denoted $|\kappa_j,\ 0 >$, 
by searching for the minimum energy of fixed configuratin of deformed s.p. orbitals. Starting from 
each basic deformed Slater determinant, a number of particle-hole excited configurations were considered
under some selection criteria (see Eq. (23) of Ref. \cite{pci1})  to keep the total numebr of basis 
states manageable. Having the deformed basis of Slater determinants chosen, one can use standard spin projection
techniques to solve the associated eigenvalues problem.
The method proved to be very accurate in $0\hbar \omega$ model spaces, such as $sd$ and $pf$ where one can
easily keep track of different deformed orbitals. The difficulties usually appear for quasi-spherical nuclei, such
as $^{56}$Ni, when special attentions has to be given to the selection of the basics states $|\kappa_j,\ 0 >$.
A simlar problem arises for the case of mixed parity valence space, such $f5pg9$ (see below), due to difficulties
in tracking fixed configurations of nucleons filling the s.p. orbitals around the level-crossing deformations.

The paper is organized as follow. Section II presents a brief outline
of the PCI formalism that was expanded in Ref. \cite{pci1}. 
The new algorithm to select the PCI basis is discussed
in section III. Section IV analyzes the efficiency of the new method 
in the case of the quasi-spherical nucleus $^{56}$Ni. Section V is devoted to 
the study of several nuclei that can be described using the mixed parity
 valence space $f5pg9$.  Conclusions and outlook are given in section VI. 

\section{The method of the Projected Configuration Interaction (PCI)}

The model Hamiltonian used in CI calculations can be written as:

\begin{eqnarray}\label{ham}
H=\sum_i e_i c_i^\dagger c_i+\sum_{i>j,k>l}V_{ijkl}c_i^\dagger c_j^\dagger c_l c_k,
\end{eqnarray}
where, $c_i^\dagger$ and $c_i$ are creation and annihilation operators of the spherical harmonic oscillator,
 $e_i$ and $V_{ijkl}$ are one-body and two-body matrix elements that can be obtained
 from effective interaction theory, such as G-Matrix plus core polarization \cite{Morten}, which 
can be further refined using the experimental data \cite{Honma02,horoi06}.

One can introduce the deformed single particle (s.p.) basis, 
which can be obtained from a constraint Hartree-Fock (HF) solution, 
or from the Nilsson s.p. Hamiltonian \cite{pci1}.
The deformed s.p. creation operator is given by the following transformation:
\begin{eqnarray}\label{spwf}
b^\dagger_k=\sum_i W_{ki}c^\dagger_i,
\end{eqnarray}
where the matrix elements $W_{ki}=\langle b_k|c_i\rangle$ are real in our calculation. 
The Slater Determinant (SD) built with the deformed single particle states is given by
\begin{eqnarray}\label{sd}
|\kappa\rangle\equiv|s,\epsilon\rangle\equiv b^\dagger_{i_1}b^\dagger_{i_2}...b^\dagger_{i_n}|\rangle,
\end{eqnarray}
where $s$ refers to the Nilsson configuration, indicating the pattern of the occupied orbits, 
and $\epsilon$ is the deformation determined by the quadrupole $\epsilon_2$, hexadecupole $\epsilon_4$ as
in Ref. \cite{pci1}, but also octupole $\epsilon_4$, etc.

The general form of the nuclear wave function is taken as a
linear combination of the projected SDs (PSDs),
\begin{eqnarray}\label{wv}
|\Psi^\sigma_{IM}\rangle=\sum_{K\kappa} f^{\sigma}_{IK\kappa}
P^I_{MK}|\kappa\rangle,
\end{eqnarray}
where $\hat P^I_{MK}$
is the angular momentum projection operator. 
The energies and the wave functions [given in terms of the coefficients
$f^\sigma_{IK\kappa}$ in Eq.(\ref{wv})] are obtained by solving
the following eigenvalue equation:

\begin{eqnarray}\label{eigen}
\sum_{K'\kappa'}(H_{K\kappa, K' \kappa'}^I-E^\sigma_IN_{K\kappa,
K' \kappa'}^I)f^\sigma_{IK\kappa'}=0,
\end{eqnarray}
where $H_{K\kappa, K' \kappa'}^I$ and $N_{K\kappa, K' \kappa'}^I$
are the matrix elements of the Hamiltonian and of the
norm, respectively
\begin{eqnarray}\label{hn}
H_{K\kappa, K' \kappa'}^I&=&\langle
\kappa|HP^I_{KK'}|\kappa'\rangle,\\ N_{K\kappa, K'
\kappa'}^I&=&\langle \kappa|P^I_{KK'}|\kappa'\rangle.
\end{eqnarray}

More details about the formalism can be found in Ref. \cite{pci1}.

\section{Choice of the PCI basis}
The analysis made in Ref. \cite{pci1} indicated that one of the most important 
problem of the PCI method is a proper selection of  the PCI basis.
As introduced in our previous work \cite{pci1}, the general structure of the PCI basis is 
\begin{eqnarray}\label{basis}
\left\{\begin{matrix}
  0{\text p}-0{\text h},  & \>  n{\text p}-n{\text h} \\
|\kappa_1,0\rangle, & |\kappa_1,j\rangle,\cdots, \\
|\kappa_2,0\rangle, & |\kappa_2,j\rangle,\cdots, \\
\hdotsfor{2}\\
|\kappa_N,0\rangle, & |\kappa_N,j\rangle,\cdots 
\end{matrix}\right\}.
\end{eqnarray}
where $|\kappa_i,0\rangle$ ($i=1,...N$) is a set of starting states of different deformations.
Assuming that we've found these $|\kappa,0\rangle$ SDs (skipping the subscript $i$ to keep notations short), 
relative $n$p-$n$h SDs, $|\kappa,j\rangle$, on top of each $|\kappa,0\rangle$
are selected using the constraint \cite{pci1}
\begin{eqnarray}\label{ecut}
\Delta E=\frac12(E_0-E_j+\sqrt{(E_0-E_j)^2+4|V|^2})\geq E_\text{cut},
\end{eqnarray}
where $E_0=\langle\kappa,0|H|\kappa,0\rangle$, $E_j=\langle\kappa,j|H|\kappa,j\rangle$,
$V=\langle\kappa,0|H|\kappa,j\rangle$ and $E_\text{cut}$ is a parameter.

The $|\kappa,0\rangle$ SDs need to be properly chosen in order to get good accuracy. 
In our previous work \cite{pci1}, we have chosen the SDs with 
the lowest unprojected expectation energy for each configuration, 
and we used the same basis for all the spins. 
That approach proved to work well for quite deformed nuclei, limiting its range of application. 
For instance, the description of $^{56}$Ni with GXPF1A \cite{horoi06} exhibits a spherical
ground state minimum which is selected as a basis SD.
This spherical SD has a good spin $I=0$,
which won't be useful if we calculate $I\neq 0$ states. 
This example suggests that a more efficient method may involve choosing different basis sets for different spins.
Another problem with this method of selecting the $|\kappa,0\rangle$ basis states is that 
for each configuration, only one SD was selected. 
Therefore some states, such as the $\beta-$vibrational states, cannot be described unless 
two or more shapes for the same configuration are artificially included outside of any algorithm.

To address these problems, we developed a new method of finding an efficient set of 
$|\kappa,0\rangle$ states. 
The deformed single particle states are generated from the Nilsson Hamiltonian shown 
in Eq. (2) of Ref. \cite{pci1}. For simplicity, we set
\begin{eqnarray}
E_i=e_i.
\end{eqnarray}
In a first step, at each deformation $\epsilon=(\epsilon_2, \epsilon_3, \epsilon_4)$, we build many Slater determinants(SDs)
 denoted by $|s,\epsilon\rangle$, where $s$ denotes a configuration of nucleons occupying the deformed
single particle orbitals.
These SDs are projected onto good angular momentum $I$, and the projected energy is calculated  
\begin{eqnarray}\label{epj}
E_\text{exp}(I,s,\epsilon)=\frac{\langle s,\epsilon|HP^I_{KK}|s,\epsilon\rangle}
{\langle s,\epsilon|P^I_{KK}|s,\epsilon\rangle}.
\end{eqnarray}

We then identify the configuration $s_a$ which has the lowest $E_\text{exp}(I,s,\epsilon)$ at 
each shape $\epsilon$.
Searching over all possible deformations $\epsilon$, we obtain the energy surface of $E_\text{exp}(I,s_a,\epsilon)$
as a function of $\epsilon$. The SD $|s_a,\epsilon_a\rangle$ which has the lowest 
$E_\text{exp}(I,s_a,\epsilon)$ is chosen as the first $|\kappa,0\rangle$ state denoted as
\begin{equation}
|\kappa_1,0\rangle=|s_a,\epsilon_a\rangle.
\end{equation}

The next step is to find the second $|\kappa,0\rangle$ state. We try all possible $|s,\epsilon\rangle$, 
and for each $|s,\epsilon\rangle$, we build the $2\times 2 $ Matrix pair $\bold{(A,B)}$,
\begin{equation*}
\bold{A}=\begin{pmatrix}
H_{11} & H_{12}\\
H_{21} & H_{22}
\end{pmatrix},
\bold{B}=\begin{pmatrix}
N_{11} & N_{12}\\
N_{21} & N_{22}
\end{pmatrix},
\end{equation*}
where 
\begin{eqnarray}
H_{ij}=\langle i|HP^I_{MK}|j\rangle,N_{ij}=\langle i|P^I_{MK}|j\rangle,\\
\text{with } |i(j)=1\rangle=|\kappa_1,0\rangle,|i(j)=2\rangle=|s,\epsilon\rangle.
\end {eqnarray}
Solving the generalized eigenvalue problem 
\begin{eqnarray}\label{ge}
\bold{A}x=\lambda\bold{B}x,
\end{eqnarray}
we get two eigenvalues, $\lambda_1$ and $\lambda_2$, and
their sum,
\begin{eqnarray}
S_2=\lambda_1^{(2)}+\lambda_2^{(2)}.
\end{eqnarray}
The SD $|s_b,\epsilon_b\rangle$ with the lowest $S_2$ is selected as the second $|\kappa,0\rangle$ denoted as
\begin{equation}
|\kappa_2,0\rangle=|s_b,\epsilon_b\rangle.
\end{equation}

The process of finding more $|\kappa,0\rangle$ SDs can be continued in a similar manner. 
Assuming that we have found the $(n-1)-$th $|\kappa,0\rangle$ SD,
$|\kappa_{n-1},0\rangle$, then $|\kappa_n,0\rangle$ is chosen as the $|s_x,\epsilon_x\rangle$, 
corresponding to the lowest $S_n$.
Here,
\begin{eqnarray}\label{sn}
S_n=\lambda_1^{(n)}+\lambda_2^{(n)}+\cdots+\lambda_{n}^{(n)}.
\end{eqnarray}
and $\lambda_1^{(n)}$,$\lambda_2^{(n)}$,..., $\lambda_n^{(n)}$ are eigenvalues of Eq.(\ref{ge}) with 
\begin{equation*}
\bold{A}=\begin{pmatrix}
H_{11} & H_{12} & \dots & H_{1n}\\
H_{21} & H_{22} & \dots & H_{2n}\\
\hdotsfor{4}\\
H_{n1} & H_{n2} & \dots & H_{nn}
\end{pmatrix},
\bold{B}=\begin{pmatrix}
N_{11} & N_{12} & \dots & N_{1n}\\
N_{21} & N_{22} & \dots & N_{2n}\\
\hdotsfor{4}\\
N_{n1} & N_{n2} & \dots & N_{nn}
\end{pmatrix},
\end{equation*}
and
\begin{eqnarray}
&&H_{ij}=\langle i|HP^I_{MK}|j\rangle,N_{ij}=\langle i|P^I_{MK}|j\rangle,\\
&&|i(j)\rangle=|\kappa_{i(j)},0\rangle, \text{ if }i(j)=1,2,\cdots,n-1;\nonumber\\
&&|i(j)\rangle=|s,\epsilon\rangle, \text{ if }i(j)=n.
\end {eqnarray}

Sometimes we may only use part of the $S_n$ sum over the lowest $\lambda_i$'s as a selection
criteria, i.e.,
\begin{eqnarray}\label{skn}
S^k_n=\lambda_1^{(n)}+\lambda_2^{(n)}+\cdots+\lambda_{k}^{(n)}, (1\leq k \leq n),
\end{eqnarray}
and $S^n_n=S_n$ by definition.

Evaluating the projected energies for all SDs takes a long time to calculate, and not all of them may
be necessary. Therefore, we enforce additional truncations.
First, the HF energy, $E_\text{HF}$, is calculated.
Next, at each shape $\epsilon$, the SDs having all particles occupying the lowest Nilsson orbits 
are considered as the 0p-0h SDs. 
All particle-hole excitations up to 4p-4h built on these 0p-0h SDs are created, 
and their expectational energies 
\begin{eqnarray}\label{exp}
E_\text{exp}(s,\epsilon)=\langle s,\epsilon|H|s,\epsilon\rangle
\end{eqnarray}
are calculated. Those SDs satisfying $E_\text{exp}(s,\epsilon)-E_\text{HF}<E_\text{expup}$, 
where $E_\text{expup}$ is a input parameter, are saved.
Finally, the projected energies $E_\text{exp}(I,s,\epsilon)$  of the saved SDs are evaluated, 
and compared with the lowest projected energy $E_\text{exp}(I,s_a,\epsilon_a)$ available.
 We keep those SDs satisfying 
$E_\text{exp}(I,s,\epsilon)-E_\text{exp}(I,s_a,\epsilon_a)<E_\text{pjup}(I)$,
 where $E_\text{pjup}(I)$ is a input parameter, and we discard the others. 
The values of the parameters $E_\text{expup}$ and $E_\text{pjup}(I)$ must
 be large enough so that the $|\kappa,0\rangle$ can be properly chosen, 
 but too large $E_\text{expup}$ and $E_\text{pjup}(I)$ value may results in wasted
computation without any improvement in accuracy.

Here we summarize the  advantages of the new method of selecting the  basis states $|\kappa,0\rangle$.
Firstly, as already mentioned, the method proposed in Ref. \cite{pci1} 
uses the same $|\kappa,0\rangle$ states for all spins, however,
certain $|\kappa,0\rangle$ states may not bring any contribution to certain spins.
The new method improves the efficiency of the PCI basis, by choosing
different $|\kappa,0\rangle$ states for different spins.

Secondly, the $|\kappa,0\rangle$ chosen by the new method may include two or more shapes
 for the same configuration $s$. Therefore, the present method explicitly
  includes the idea imbedded in the Generator Coordinate Method (GCM) \cite{HW53},
 and may be used to describe some collective vibrations, such as
 the $\beta-$vibration.

Thirdly, there are no limitations on the $K$ values. In Ref. \cite{pci1}, we only selected
basis states with relatively small-$K$ values to keep the basis dimensions manageable. 
This limitation could be a problem in the case of the high spin states, 
for which the high-$K$ configurations could be close to the yrast line. 
The new method described here selects basis states with all possible SDs satisfying $|K|\leq I$.

Finally, and perhaps more importantly, the drawback of getting large overlaps between different basis SDs, 
which is typical for an uncorrelated selections of the basis, is avoided by construction in the
new method. Getting basis states with large overlaps leads to many spurious states due to the
zero eigenmodes of the norm matrix. The significance of these zero modes is that some of the basis 
states that exhibit large overlaps bring insignificant contribution to the solutions of the Hill-Wheeler 
Eq. (5), while unnecessarily increasing the dimensions of the problem.
 Those useless SDs can be automatically
filtered out by the present method because the overlap problem has been
fully considered step-by-step when the generalized eigenvalue equations (\ref{ge}) are solved.

\section{Calculations of $^{56}$Ni}

\begin{figure}
\centering
\includegraphics[width=3.5in]{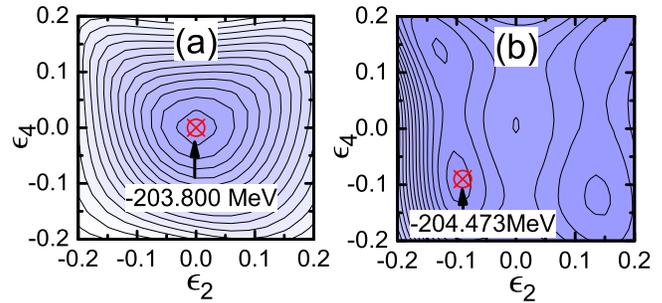}
\caption{(Color online) Unprojected energy surface (left panel) and the projected
energy surface with $I=0$ (right panel) for the ground state of $^{56}$Ni
 with the GXPF1A interaction. The lowest energy is marked by '$\oplus$'} 
\label{Ni56pes}
\end{figure}

Using the new method, we recalculated the nuclei $^{56}$Ni with GXPF1A interaction\cite{horoi06}. 
Let's first consider the case of $I=0$. Both $\epsilon_2$ and $\epsilon_4$ span the interval 
from $-0.45$ to $0.45$ in steps of $0.03$.
The first basis state, $|\kappa_1,0\rangle$, having the lowest projected energy, corresponds to $I=0$. 
In Fig. \ref{Ni56pes},
 the surfaces of the $E_\text{exp}(I,s_a,\epsilon)$ (See Eq. (\ref{epj})) and $E_\text{exp}(s_a,\epsilon)$ 
(See Eq. (\ref{exp}))
 are plotted as functions of $\epsilon_2$ and $\epsilon_4$. Our calculation shows that the 
 configuration $s_a$ has all 
16 valence particles in $^{56}$Ni occupying the orbits coming from the $1f_{7/2}$ subshell. 
The unprojected minimum, Fig. (1)$(a)$ is at $-203.800$ MeV, and its shape is spherical, 
consistent with the HF result. 
However, the projected energy surface presents a quite different picture. 
There are four minima around the spherical shape, 
the lowest one has $-204.473$ MeV and a small oblate deformation, $\epsilon_2=-0.09$ and $\epsilon_4=-0.09$. 
This energy is $673$ keV lower than the HF energy, and $1.236$ MeV above the exact CI  ground state energy of
$-205.709$ MeV. 
This $|\kappa, 0\rangle$ state is a good candidate for the ground state.

The second basis state, $|\kappa_2,0\rangle$, has the same configuration as the first one,
 but a different shape characterized by $\epsilon_2=-0.24$ and $\epsilon_4=-0.15$. 
Comparing with Fig. \ref{Ni56pes}, this shape is quite different from any of the remaining 3 minima. 
The reason is that the SDs at those 4 minima are highly overlapping each other after
 the angular momentum projection. Once the lowest one is picked up,
 the others will automatically be filtered out by the present method. The
$|\kappa_2,0\rangle$ corresponds to the first excited $0^+$ state, 
which might be called a $\beta-$vibrational state.

\begin{figure}
\centering
\includegraphics[height=3.0in]{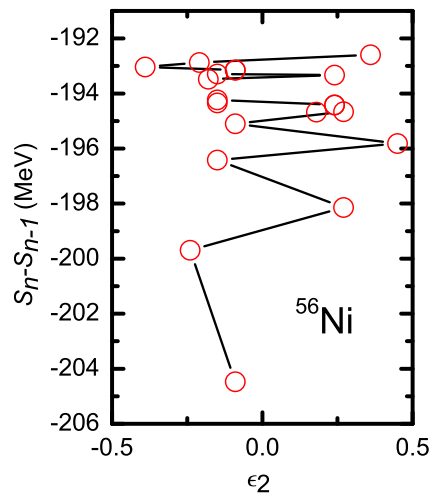}
\caption{(Color online) $S_i-S_{i-1}$ values of $|\kappa_i,0\rangle$ at $I=0$ for $^{56}$Ni as a function of $\epsilon_2$. 
$\epsilon_4$ are included in the calculation. $|\kappa_1,0\rangle$ is the lowest one.  } 
\label{Ni56k0e}
\end{figure}

The third basis state, $|\kappa_3,0\rangle$, has a prolate shape with $\epsilon_2=0.27$ and $\epsilon_4=0.06$.
Its configuration can be obtained starting from $|\kappa_1,0\rangle$, but with 4 particles 
jumping from the $|\Omega|=7/2$ ($1f_{7/2}$) orbits
 to the $|\Omega|=1/2$ ($2p_{3/2}$) orbits. $|\kappa_3,0\rangle$ can generate
a deformed rotational band, which has been observed in experiments \cite{Ni99}. 
Here, we only create the band head, which is the third $0^+$ state.
Informations about higher $|\kappa,0\rangle$ SDs are shown in Fig. \ref{Ni56k0e}.
 The value $S_i-S_{i-1}$ indicates the energy position of each $|\kappa_i,0\rangle$ state.

Once we have selected the $|\kappa,0\rangle$ SDs, we perform the PCI calculations. 
There are two parameters used in PCI:
one is the number of $|\kappa,0\rangle$ SDs, $n$, and the other is the $E_\text{cut}$
used in Eq. (9) to select the number of particle-hole excitations on top of 
each $|\kappa,0\rangle$ \cite{pci1}. 

\begin{figure}
\centering
\includegraphics[width=3.5in]{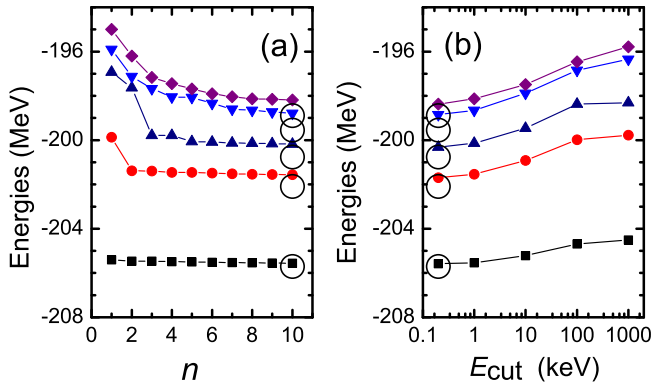}
\caption{(Color online) Left panel, $I=0$ PCI energies for $^{56}$Ni as functions of $n$ 
with $E_\text{cut}=1$ keV. 
Right panel, $I=0$ PCI energies for $^{56}$Ni as functions of $E_\text{cut}=1$ with $n=8$.
Full CI results are also shown as the open circles by comparison.} 
\label{Ni56n}
\end{figure}

It is interesting to study how many $|\kappa,0\rangle$ are needed to describe the low-lying states. 
Fig. (3)$(a)$ shows the PCI energies   as functions of 
$n$ for $E_\text{cut}=1$ keV.
For $n=1$ the  PCI dimension is only 180, yet the first $0^+$ PCI energy is $-205.409$ MeV,
 only 300 keV above the exact full CI value.
$|\kappa_2,0\rangle$ must be included to describe the second $0^+$ state, and
$|\kappa_3,0\rangle$ reproduces the third $0^+$ state. To accurately describe more excited states 
more $|\kappa,0\rangle$ SDs are need. For the lowest 5 states in $^{56}$Ni a good approximation can 
been achieved starting with $n=7$.
If one keeps on increasing $n$, then those 5 energies will become closer and closer to the CI values. 
For instance,
with $n=15$, the the ground state PCI energy becomes $-205.603$ MeV, just 100 keV above the exact value.

\begin{figure}
\centering
\includegraphics[width=3.0in]{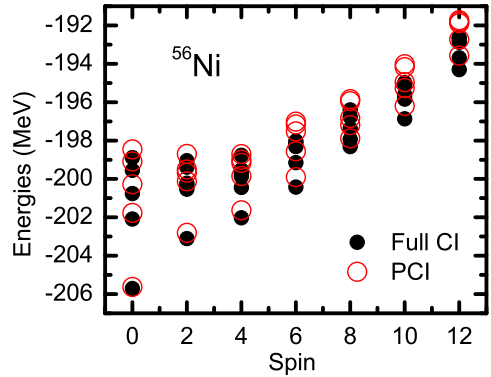}
\caption{(Color online)The lowest 5 energies at each spin for $^{56}$Ni
 calculated using PCI (open circles) and full CI (filled circles). } 
\label{Ni56e}
\end{figure}

The PCI energies are also affected by the $E_\text{cut}$ parameter. For $E_\text{cut}=1000$ keV 
no particle-hole excited SDs are included, and the PCI dimension is the same as the number of
$|\kappa,0\rangle$ SDs included, $n=8$ in this case. Therefore the PCI 
energies are exactly the values of $\lambda_i$ in Eq.
(\ref{sn}). More particle-hole excited SDs can be included by reducing 
the value of $E_\text{cut}$.
For example, by decreasing from $E_\text{cut}=1000$ keV to $E_\text{cut}=1$ keV, 
the PCI energies drop $\sim 1.0 \div 2.3$ MeV
for the lowest 5 states, and become close to the full CI values. 
By decreasing from $E_\text{cut}=1$ keV to $E_\text{cut}=0.2$ keV the energy drop 
becomes slower, and is around $\sim 100 \div 200$ keV.
The PCI energies for $I\neq 0$ states are also calculated, and shown in Fig. \ref{Ni56e}, 
the number of $|\kappa,0\rangle$ is $n=15$ and $E_\text{cut}=1$ keV. One can observe good agreements 
between the PCI and the CI results, including the states in the rotational band starting at about 5 MeV.

\section{calculations in $f5pg9$ valence space}

We have also extended our calculations to the $f5pg9$ valence space, which 
includes the $1f_{5/2}$, $2p_{3/2}$, $2p_{1/2}$, and the $1g_{9/2}$ spherical shells. 
$1g_{9/2}$ orbital has positive parity, and the other $fp$ orbitals have negative parity.
Therefore, SDs with both parities can be built for any number of nucleons. 
The positive parity SDs are those with even number particles occupying the $fp$ orbitals, 
while the negative parity SDs have odd number particles occupy the $fp$ orbitals.
The angular momentum projection does not change the parity. The parity of the projected states
 remain the same as that of the original SDs. Therefore, one can split the PCI basis 
 into the positive parity part, $I^+$, and the negative parity part, $I^-$, at each spin $I$. 
With the present method, different $|\kappa, 0\rangle$ 
SDs can be generated separately for the $I^+$ basis and $I^-$ basis.

\begin{figure}
\centering
\includegraphics[width=3.5in]{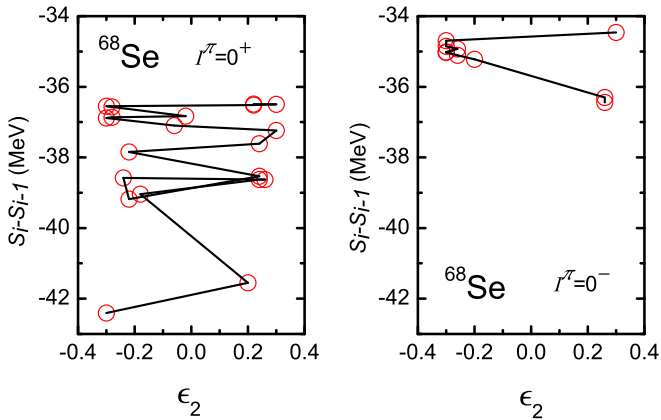}
\caption{(Color online)  $S_i-S_{i-1}$ values of $|\kappa_i,0\rangle$ for 
 $I^\pi=0^+$ (left) and $I^\pi=0^-$ (right) in $^{68}$Se as a function of $\epsilon_2$. 
$\epsilon_4$ are included in the calculation. } 
\label{Se68k0}
\end{figure}

The interaction for $f5pg9$ shell space was taken from Ref. \cite{Ka04}. 
It includes, besides the usual qqudrupole, hexadecupole, and pairing terms,
octupole contributions and monopole corrections.
In all cases, 20 $|\kappa,0\rangle$ SDs are taken for each $I^\pi$ basis 
and $E_\text{cut}$ was fixed to 1 keV.
The first example we analyze is the N=Z nucleus $^{68}$Se, which is known to be
deformed with competing oblate and prolate deformations. 
The energies of the 20 $|\kappa,0\rangle$ SDs for both $I^\pi=0^+$ basis
 and $I^\pi=0^-$ basis are  shown in Fig. \ref{Se68k0}. 

\begin{figure}
\centering
\includegraphics[width=2.0in]{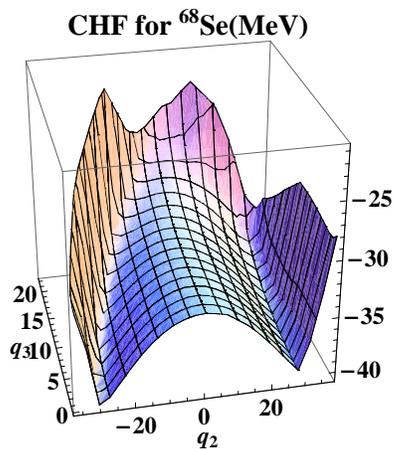}
\caption{(Color online)  Energy surface provided by Constrained Hartree-Fock (CHF) calculations 
as a function of 
$q_{2}=\sqrt {\frac{16\pi}{5}}\left (\frac{r}{b}\right )^2 Y_{20}$ and 
$q_{3}=\left (\frac{r}{b}\right )^3 Y_{30}$. ($b$ is the harmonic oscilater length) }
\label{se68_chf}
\end{figure}

It is known that $^{68}$Se nucleus exhibits shape coexistence features. 
The constrained HF calculations of Ref. \cite{Ka04} as well as our results in 
Fig. \ref{se68_chf} show
that there are two minima. Both of them are axially and reflection symmetric. 
The lowest minimum has $-40.718$ MeV and oblate shape,
and the second one has $-39.956$ MeV and prolate shape.
 It is interesting that the results of our new method presents the same picture
as one can observe in the left panel of Fig. \ref{Se68k0}, where the lowest
$|\kappa_1,0\rangle$ is oblate with $S_1=E_\text{exp}(0,s_a,\epsilon_a)=-42.405$ MeV
and $|\kappa_2,0\rangle$ is prolate with $S_2-S_1=-41.549$ MeV. 
However, both energies are about $1.6$ MeV lower than those obtained by an HF procedure, 
due to the angular momentum projection.

For the $0^-$ basis the lowest $|\kappa,0\rangle$ SD lies at $-36.44$MeV,
which is relatively high (see Fig.\ref{Se68k0}), and has a prolate shape. 
The corresponding configuration is the same as that of the second HF minimum,
except that one particle was excited from the $\Omega=3/2(p_{3/2})$ orbital to 
the $\Omega=3/2(g_{9/2})$ orbital.
Because $^{68}$Se has $N=Z$, this excited particle can be either a neutron or a proton,
and therefore there are two different $|\kappa,0\rangle$ SDs with the same shape
and the same low energy. Similar cases appear for other $|\kappa,0\rangle$ SDs. 
Therefore, in Fig. \ref{Se68k0}$(b)$
 each symbol corresponds to two different $|\kappa,0\rangle$ SDs. The position of the 
  second lowest symbol is only about $130$ keV above the lowest one, and its configuration is 
  similar to the lowest one, but with the odd particle 
   excited from the $\Omega=1/2(p_{3/2})$ orbital to the $\Omega=1/2(g_{9/2})$ orbital.

\begin{figure}
\centering
\includegraphics[width=3.5in]{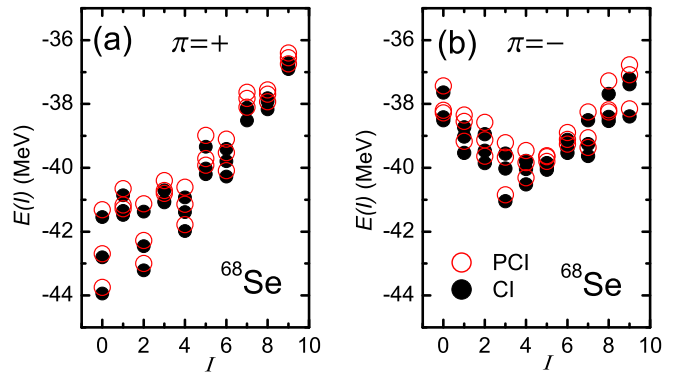}
\caption{(Color online) The lowest 3 energies at each spin/parity for $^{68}$Se
 calculated using PCI (open circles) and full CI (filled circles). } 
\label{Se68pci}
\end{figure}

Using the $|\kappa,0\rangle$ states described in Fig. \ref{Se68k0}, 
one can calculate the PCI energies for the $0^+$ and $0^-$ 
states, which turned out to be very close to the CI results.
For  $I\neq 0$, similar good results have also
been achieved, as shown in Fig. \ref{Se68pci}. 

\begin{figure}
\centering
\includegraphics[width=3.5in]{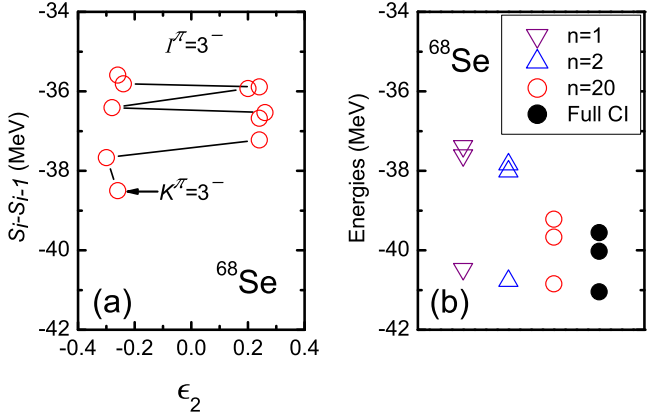}
\caption{(Color online) (a), $S_i-S_{i-1}$ values of $|\kappa,0\rangle$ at 
 $I^\pi=3^-$ in $^{68}$Se. (b), PCI energies (Open Symbols) 
 and CI energies (Fill Circles) at $I^\pi=3^-$ in $^{68}$Se.
  The open Symbols from left to right refer to PCI calculations with $n=1,2$ and $20$.}
\label{Se68i3n}
\end{figure}

As indicated in the above discussions, the PCI method not only provides a good approximation for the CI results, 
but it is also a convenient tool to gain some insight into the physics of the nuclear states. 
One interesting example is the lowest state in Fig. \ref{Se68pci}(b), which is a $3^-$ state. 
As shown in Fig. \ref{Se68i3n},
the (lowest) $|\kappa_1,0\rangle$ SD has $K^{\pi}=3^-$ and oblate deformation. 
The configuration of this $|\kappa_1,0\rangle$
is the same as the oblate HF minimum, except that one particle was excited
from the $\Omega=3/2(p_{3/2})$ orbital to the $\Omega=9/2(g_{9/2})$ orbital to form a $K^\pi=3^-$ SD. 
The second SD,  $|\kappa_2,0\rangle$, has the same energy and the same shape as
$|\kappa_1,0\rangle$ because  $N=Z$, and due to the isospin symmetry of the adopted Hamiltonian. 
If only the particle-hole excitations built on $|\kappa_1,0\rangle$ are 
included, one obtains a PCI energy of $-40.469$ MeV. 
If the $|\kappa_2,0\rangle$ SD is further included, the PCI energy drops $300$ keV to $-40.769$ MeV, 
This energy is only $300$ keV above the exact value of $-41.043$ MeV.
However, the PCI energy for $n=20$ is $-40.843$ MeV, only $70$ keV lower than what one can obtain
with $n=2$. Therefore, it is clear that the lowest  $3^-$ state has mostly contributions from the 
lowest 2 oblate $K^\pi=3^-$ SDs, i.e. $|\kappa_1,0\rangle$ and $|\kappa_2,0\rangle$.

\begin{figure}
\centering
\includegraphics[width=3.5in]{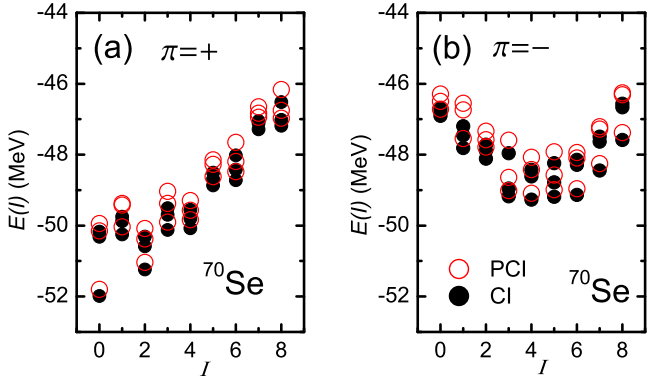}
\caption{(Color online) The same as Fig. \ref{Se68pci} but for $^{70}$Se. } 
\label{Se70pci}
\end{figure}

\begin{table}
\caption{\label{dim} PCI Dimensions compared with those of full CI for $^{70}$Se}
\begin{ruledtabular}
\begin{center}
\begin{tabular}{ccccc}

Spin  & \multicolumn{2}{p{2.5cm}} {\centering $\pi=+$}&\multicolumn{2}{p{2.5cm}} {\centering $\pi=-$}\\
\cline{2-3}\cline{4-5}
($I$)& PCI &CI &PCI &CI\\
\hline
0 & 3665 & $ 6.7\times 10^6 $ & 4497  & $ 6.7\times 10^6 $\\
1 & 4735 & $ 2.0\times 10^7 $ & 4431  & $ 2.0\times 10^7 $ \\
2 & 4369 & $ 3.2\times 10^7 $ & 4284  & $ 3.2\times 10^7 $ \\
3 & 4799 & $ 4.2\times 10^7 $ & 4778  & $ 4.2\times 10^7 $ \\
4 & 4384 & $ 5.0\times 10^7 $ & 4476  & $ 5.0\times 10^7 $ \\
5 & 4714 & $ 5.5\times 10^7 $ & 4284  & $ 5.5\times 10^7 $ \\
6 & 4246 & $ 5.8\times 10^7 $ & 4636  & $ 5.8\times 10^7 $ \\
7 & 4505 & $ 5.9\times 10^7 $ & 4159  & $ 5.9\times 10^7 $ \\
8 & 4125 & $ 5.7\times 10^7 $ & 4056  & $ 5.7\times 10^7 $ 
\end{tabular}
\end{center}
\end{ruledtabular}
\end{table}

We have also calculated states of both parities in $^{70}$Se. The results are shown in Fig. \ref{Se70pci}. 
Once again, the PCI results are very close to those of full CI for both positive parity
 and negative parity for a wide range of spin values. However, much smaller dimensions of the PCI
matrices are necessary. 
The PCI dimensions correponding to the $I^\pi$ calculations in  Fig. \ref{Se70pci}
  are compared in Table \ref{dim} with the full coupled-I CI dimensions. 
 The PCI dimensions are small fractions, roughly $10^{-4}$, of the full CI dimensions.
 As is well known, the most serious problem with full CI method is the explosion of
 the dimensions as the number of the single-particle valence states, and/or number of valence nucleons. 
 However, this problem seems to be less of an issue for the PCI method. 
The total PCI dimension can be estimated as
the product of two numbers,
 $ n\times m $, where $n$ is the number of $ |\kappa,0\rangle $ states and $m$ is
 the number of particle-hole excitations selected by $E_\text{cut}$ in Eq. (\ref{ecut}). 
 Our investigations indicate that $n$ is related to
  how many low-lying states of a given spin one wants to accurately describe.
  For instance, if we are interested in only the yrast state, 
  quite often a good approximations can be obtained with $n=1$ or 2.
  $n=20$ seems to be enough to describe the lowest 3-5 states of each $I^\pi$ in the present calculations. 
  As regarding $m$, the $|\kappa,j\rangle$ SDs are limited to 1p-1h and 2p-2h excitations 
  according to Eq. (\ref{ecut}). 
  Note that $|\kappa,j\rangle$ has the same $K^\pi$ as that of $|\kappa,0\rangle$. 
  As the parameter $E_\text{cut}$ is enforced via Eq. (\ref{ecut}), $m$
 can be significantly reduced. For instance,
  in the case of $I^\pi=0^+$ for $^{70}$Se, $m_\text{2p2h}=801$ for $|\kappa_1,0\rangle$,
   but only 138 SDs were finally included if $E_\text{cut}=1$ keV.
 
The exploding CI dimensions has as a consequence a rapid increase of 
the computing time necessary for full CI calculation.
Using the modern coupled-I code NuShellX \cite {nushellx}, the full CI calculation 
of all states in Fig. \ref{Se70pci} 
could take almost one year when only
one processor is used. The calculation of the lowest 3 states of each $I^\pi$ in $^{70}$Se would take 
in average about 20 days.
For the same calculation, PCI takes around 5 hours for each $I^\pi$.
The main computational workload in PCI is related to the calculation of
the dense matrices $H$, and $N$ in Eq. (\ref{hn}).  
It should be metioned that extra time is needed to 
to search for the optimized set of $|\kappa,0\rangle$ SDs.
The computing time can be affected by: (1) the number of mesh points used for the shape paramters;
 (2) the values of the parameters $E_\text{expup}$ and $E_\text{pjup}(I^\pi)$ that decides how many SDs are considered in the optimization process;
 (3) the total number $n$ of $|\kappa,0\rangle$ basis states selected;
 (4) the value of $E_\text{cut}$. 
For example, in the calculation of $I^\pi=0^+$ in $^{70}$Se,
both $\epsilon_2$ and $\epsilon_4$ run from $-0.3$ to $0.3$
 in steps of $0.02$, $E_\text{expup}=7$ MeV and $E_\text{pjup}(I^\pi=0^+)=5$ MeV, and $E_\text{cut}=1$.
Under these conditions, it will takes about 10 hours to obtain 20 $|\kappa,0\rangle$ SDs using one processor. 
For other $I^\pi$, the computational time ranges from few hours to 1 or 2 days. 
However, the total time for a PCI calculations is at least 10 times shorter
 than that of the corresponding full CI calculation for the case of $^{70}$Se.

\begin{figure}
\centering
\includegraphics[width=3.5in]{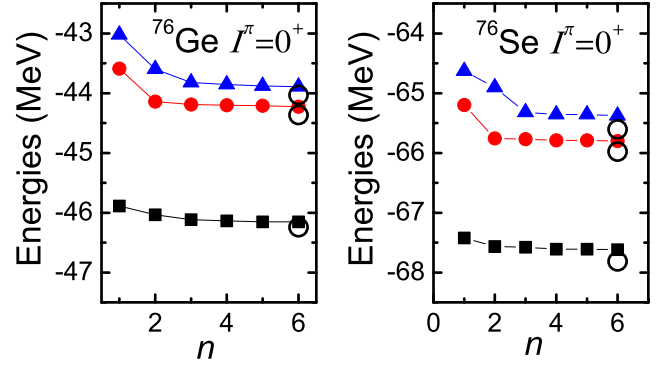}
\caption{(Color online) Lowest 3 $0^+$ energies of $^{76}$Ge and $^{76}$Se calculated by PCI (filled symbols)
 and full CI (open circles)} 
\label{dbd76}
\end{figure}

Finally, we used the new PCI method to calculate the low lying $0^+$ states in $^{76}$Ge and $^{76}$Se.
The nuclear structure of these two nuclei is relevant for the double beta decay (DBD) process of $^{76}$Ge.
DBD is one of the most actively investigated nuclear physics problem, which may reveal new physics beyond 
the Standard Model, including the absolute values of the neutrino masses. 
Full CI calculations \cite{Ca96,Ho07,Ca08} of the 2-neutrino and neutrinoless DBD matrix elements have been 
carried out for some DBD nuclei up to $^{136}$Xe. However, for heavier DBD nuclei  $^{150}$Nd and $^{238}$U, 
the huge CI dimensions make the full CI calculation unmanageable. 
PCI can take full advantage of the deformation, and an
efficient truncation could be obtained for well deformed nuclei, such as $^{150}$Nd and $^{238}$U. 
As a first inroad into this problem, the low-lying $0^+$ 
states $^{76}$Ge and $^{76}$Se are calculated using the present
version of the PCI, and  are compared with full CI results in Fig. \ref{dbd76}.
Using only 6 $|\kappa,0\rangle$ SDs ($n=6$) for each nucleus, 
the PCI dimensions are 561 and 647 for $^{76}$Ge and $^{76}$Se, respectively.
The calculated PCI energy of the lowest $0^+$ state for $^{76}$Se is 
 200 keV higher than the exact value, and only 86 keV higher for $^{76}$Ge. 
In addition, good approximations for the excited $0^+$ states have also been reached. 
Given these encouraging results, one would hope that PCI calculations could be successfully performed
for the heavy deformed DBD nuclei, such as $^{150}$Nd and $^{238}$U, in a not so distant future.

\section{Conclusions and Outlook}

In this article we propose a newly improved algorithm of selecting the basis of Slater determinants
that can be used with the Projected Configuration Interaction method introduced in Ref. \cite{pci1}.
The new algorithm depends on a number of parameters that can be used to fine tune its efficiency. Its
main advantages over the original method of selecting of the basis are summarized at the end of
Section III.

We used the new algorithm to revisit the calculation of $^{56}$Ni, quasi-shperical nucleus that has 
a relatively low-lying rotational band. We were able to calculate its low-lying states very efficiently, 
and with good accuracy, gaining also insight into the physics of these states. 
We have also use the new method to analyze some Se and Ge isotopes in the $f5pg9$ valence space.
Both natural and unnatural parities can be accurately described for these nuclei, even for cases with
pronounced competing deformations, such as $^{70}$Se and $^{70}$Se. The PCI dimensions are significantly
lower the corresponding CI dimensions, as well as the corresponding computational effort.
In addition, in most cases, the low-lying projected basis states can provide some physical 
insight into the structure of the low-lying states.
Finally, we calculated with the new method the low-lying $0^+$ states in $^{76}$Ge and $^{76}$Se that
are relevant for the double beta decay of $^{76}$Ge. The hope is that this method could be used some day to 
study the double beta decay of the strongly deformed $^{150}$Nd and $^{238}$U.

Further improvements to the PCI method will include the extension of the formalism developed in
Ref. \cite{pci1} to calculate electromagnetic transition probabilities. The new method uses different bases
for different spins, which introduces additional complications. Other observables, such as 
spectroscopic amplitudes and DBD matrix elements have to be worked out.
Further improvement of the basis may be achieved for some cases that exhibit significant octupole
deformation, which will require full projection on good parity.

\vspace*{-8mm}

\begin{acknowledgments}

The authors acknowledge support from the DOE Grant No. DE-FC02-09ER41584.
M.H. acknowledges support from NSF Grant No. PHY-0758099.
Z.G. acknowledges the NSF of China Contract Nos. 10775182, 10435010 and 10475115.
\end{acknowledgments}
\vspace*{-6mm}


\end{document}